\documentclass[12pt]{article}
\usepackage{amsmath,amssymb,graphicx,epsfig}
\def\bea{\begin{eqnarray}}
\def\eea{\end{eqnarray}}
\def\be{\begin{equation}}
\def\ee{\end{equation}}
\def\ba{\begin{array}}
\def\ea{\end{array}}

\font\tenrsfs=rsfs10
\font\sevenrsfs=rsfs7
\font\fiversfs=rsfs5
\newfam\rsfsfam
\textfont\rsfsfam=\tenrsfs
\scriptfont\rsfsfam=\sevenrsfs
\scriptscriptfont\rsfsfam=\fiversfs
\def\mathscr#1{{\fam\rsfsfam\relax#1}}

\begin{document}
\thispagestyle{empty}
\begin{center}
\begin{small}
\hfill NEIP/06-02
\end{small}
\begin{center}

\vspace{1.5cm}

{\LARGE \bf Locally stable non-supersymmetric\\[3mm] Minkowski vacua in supergravity}

\end{center}

\vspace{1cm}

{\large \bf M.~G\'omez--Reino} {\large and} {\large \bf C.~A.~Scrucca} \\[2mm]

\vspace{7mm}

{\em Institut de Physique, Universit\'e de Neuch\^atel,\\ 
Rue Breguet 1, CH-2000 Neuch\^atel, Switzerland}
\vspace{.2cm}

\end{center}

\vspace{0.4cm}
\centerline{\bf Abstract}
\vspace{-0.1cm}
\begin{quote}

We perform a general study about the existence of non-supersymmetric minima with 
vanishing cosmological constant in supergravity models involving only chiral superfields. 
We study the conditions under which the matrix of second derivatives of the scalar potential 
is positive definite. We show that there exist very simple and strong necessary conditions 
for stability that constrain the K\"ahler curvature and the ratios of the 
supersymmetry-breaking auxiliary fields defining the Goldstino direction.
We then derive more explicitly the implications of these constraints in the case where the 
K\"ahler potential for the supersymmetry-breaking fields is separable into a sum of terms 
for each of the fields. We also discuss the implications of our general results on the dynamics 
of moduli fields arising in string compactifications and on the relative sizes of their auxiliary fields, 
which are relevant for the soft terms of matter fields. We finally comment on how the idea of uplifting 
a supersymmetric AdS vacuum fits into our general study.

\vspace{5pt}
\end{quote}

\newpage

\renewcommand{\theequation}{\thesection.\arabic{equation}}

\section{Introduction}
\setcounter{equation}{0}

Supergravity models aiming to provide a viable extension of the standard model must
have certain characteristic in order to be compatible with
present--day experiments. In particular they must lead to a tiny
cosmological constant and a moderately large effective supersymmetry breaking
scale. The general framework commonly considered involves a visible
sector containing superfields $Q_a$ describing ordinary matter and gauge particles
and their superpartners, and a hidden sector containing additional superfields $\Phi_i$. 
Supersymmetry breaking then occurs spontaneously in the hidden sector and is
transmitted to the visible sector only through gravitational interactions
\cite{gravmed,gravmedunif}. In the visible sector the net effect of this breaking can be parametrized by a
finite number of soft breaking terms, the form of these soft terms being the central issue
concerning the phenomenology of these scenarios \cite{softterms}. The dynamics of
the hidden sector induce non-vanishing vacuum expectation values for its scalar
and auxiliary fields. The soft terms in the visible sector
arise then from the dependence of the wave-function factors and couplings of the $Q_a$'s
on the hidden sector superfields $\Phi_i$ (see for instance \cite{soni,bimsoft}).
Therefore the dynamics of the hidden sector control two crucial aspects of the theory:
the size of the cosmological constant and the relative sizes of the different contributions to 
soft terms. In the lack of a theoretical framework explaining in a natural way the 
characteristics that these quantities must have, one must then pragmatically impose these 
as constraints on the theory through a tuning of parameters. In this sense a general characterization 
of the conditions under which such a sector can stabilize all its fields with masses of the order of the 
supersymmetry-breaking scale and, at the same time, produce a negligible cosmological constant 
would therefore be very interesting. 

The aim of this paper is to explore in a general and systematic way the conditions for the existence of
non-supersymmetric extrema of the scalar potential of supergravity theories fulfilling two basic
properties: i) they are locally stable and ii) they lead to a cosmological constant that is tuned to zero, 
that is, to a Minkowski space-time.
The first property guarantees that the squared masses of all the fluctuations around such a vacuum
are positive, but it does not exclude the existence of additional locally stable vacua with lower energy,
that could lead to a tunneling instability. For example, there might
exist additional supersymmetric AdS extrema (which are always stable \cite{Breitenlohner:1982bm}). 
The tunneling to such vacua is however possible only under certain
conditions \cite{Coleman:1980aw}. The second property guarantees that the small value required for the
cosmological constant can be adjusted order-by-order in the expansion defining the effective theory.

The paper is organized as follows: In Section 2 we outline the general strategy we will follow to find 
necessary conditions for the stability of non-supersymmetric Minkowski extrema of the potential in a 
general supergravity theory with $n$ chiral superfields. In Section 3 we consider the special case of 
theories with a separable K\"ahler potential and  we compute the matrix of second derivatives of the 
potential. We also write the form of the mass matrix and we derive the necessary conditions 
for this matrix to be positive definite. In Section 4 and Section 5 we study in detail the cases with one 
and two chiral superfields respectively, and we derive the exact form of the necessary condition for 
stability of the vacuum, which depends only on the K\"ahler potential. We also derive the bounds on 
the values that the auxiliary fields can take. In Section 6 we generalize this to the $n$ field case. 
In Section 7 we apply our results to the particularly interesting case of moduli fields arising from string 
compactifications, and in Section 8 we examine how the idea of uplifting an AdS supersymmetric vacuum 
fits into our study. Finally in Section 9 we summarize our results.

\section{Non-supersymmetric Minkowski minima}
\setcounter{equation}{0}

The Lagrangian of the most general supergravity theory with $n$ chiral superfields is entirely defined
by a single arbitrary real function $G$ depending on the corresponding chiral superfields $\Phi_i$ and
their conjugates $\Phi_i^\dagger$, as well as on its derivatives \cite{sugra}. The function $G$ can be 
written in terms of a real K\"ahler potential $K$ and a holomorphic superpotential $W$ in the following 
way:
\be
G(\Phi_i,\Phi_i^\dagger) = K(\Phi_i,\Phi_i^\dagger) + \log W(\Phi_i) +
\log \bar W(\Phi_i^\dagger) \,.
\ee
The quantities $K$ and $W$ are however defined only up to K\"ahler transformations acting as
$K \to K + f + \bar f$ and $W \to W + f$, $f$ being an arbitrary holomorphic function of the superfields,
which leave the function $G$ invariant by construction. We find it more convenient for our purposes
to work with the function $G$.

The part of the supergravity Lagrangian that will be relevant for our analysis is the scalar potential,
which has the following simple form\footnote{We use Planck units where $M_{\rm P}=1$, and the standard
notation $G_i \equiv \partial G/\partial {\Phi_i}$, $G_{\bar i} \equiv \partial G/\partial{\Phi_i^\dagger}$,
etc ..., the indices being lowered and raised with the K\"ahler metric $G_{i\bar j}$ and its inverse 
$G^{i \bar j}$.}:
\be
\label{genpot}
V = e^G \left(G^{i\bar j}G_i G_{\bar j}-3\right) \,.
\ee
In order to find local non-supersymmetric Minkowski minima of the potential (\ref{genpot}), one should proceed
as follows: First find the points that satisfy the stationarity conditions $\langle V_I \rangle =0$ (for $I=i,\bar i$
and $i=1,\cdots,n$). Then impose the Minkowski condition $\langle V \rangle=0$ and require, at the same time, 
that $\langle W \rangle \neq 0$ so that supersymmetry is broken. And finally verify that the matrix 
$\langle V_{IJ} \rangle$ (with $I=i,\bar i$, $J=j,\bar j$ and $i,j=1,\cdots,n$) of second derivatives of the potential 
(that is, the Hessian matrix) is positive definite.

The first part of this program can be carried out in full generality by using the tools of
K\"ahler geometry, which are based on the fact that the metric $G_{i \bar j}$ is obtained as the second derivative
of the potential $G$ \cite{supertrace} (see also \cite{Ferrara:1994kg,softD}). The only two types of non-vanishing Christoffel 
symbols entering in the covariant derivatives are those with only holomorphic or anti-holomorphic indices, namely 
$\Gamma_{i j}^k = G_{i j \bar l} G^{\bar l k}$ and $\Gamma_{\bar i \bar j}^{\bar k} = G_{\bar i \bar j l} G^{l \bar k}$. 
The Minkowski condition\footnote{In what follows, we will omit the expectation value symbol $\langle \dots \rangle$ 
from all the equations, but it should be understood that they are evaluated on the vacuum, unless otherwise noted.} 
$V = 0$ following from (\ref{genpot}) implies that:
\be
\label{3}
G^k G_k = 3 \,.
\ee
Using this Minkowski condition the stationarity conditions can be equivalently rewritten as $\nabla_i V = 0$ and 
they imply:
\be\label{4}
G_i + G^k \nabla_i G_k = 0 \,.
\ee
Finally, the second derivatives of the potential can be computed as well by using covariant
derivatives, since the extra connection terms vanish by the Minkowski and stationarity conditions.
There are two different $n$-dimensional blocks, $V_{i \bar j} = \nabla_i \nabla_{\bar j} V$ and
$V_{i j} = \nabla_i \nabla_j V$, and after a straightforward computation these are found to be given 
by the following expressions:
\bea\label{vij}
\begin{array}{lll}
V_{i \bar j} \!\!\!&=&\!\!\! e^G \Big(G_{i \bar j} + \nabla_i G_k \nabla_{\bar
j} G^k - R_{i \bar j p \bar q} G^p G^{\bar q}\Big) \,, \smallskip\ \\
V_{i j} \!\!\!&=&\!\!\! \displaystyle{e^G \Big(\nabla_i G_j + \nabla_j G_i + \frac 12 G^k \big\{\nabla_i,\nabla_j \big\}G_k}\Big) \,. \\
\end{array}
\eea
The whole $2n$-dimensional matrix of second derivatives is then given by
\begin{equation}
V_{IJ} = \left(
\begin{matrix}
V_{i \bar j} &  V_{ij} \smallskip \cr
V_{\bar i \bar j} & V_{\bar i j}
\end{matrix}
\right) \,.
\label{VIJ}
\end{equation}
The conditions under which this $2n$-dimensional matrix is positive definite are however difficult 
to work out in general, the only way being to study in full detail the behavior of all the $2n$ eigenvalues. 
More precisely said, they do not seem to translate into simple necessary and sufficient conditions on the
potential $G$ specifying the theory.

The main aim of this paper is to try to deduce some simple necessary conditions for the Hessian matrix 
(\ref{VIJ}) to be positive definite. In order to do so the crucial point that we will exploit 
is the fact that the requirement for a matrix to be positive definite is equivalent to the requirement that all its 
upper-left subdeterminants are positive, that is, to the requirement that all its upper-left submatrices are positive 
definite. In our case, this implies in particular that the $n$-dimensional submatrix $V_{i \bar j}$ should be positive definite:
\be
V_{i \bar j} \;\, \mbox{positive definite}\,.
\label{pos}
\ee
This means by definition that the quadratic form $V_{i \bar j} z^i {\bar z}^{\bar j}$ should be positive
for any choice of non-null complex vector $z^i$. Our strategy will be to look for a suitable 
vector $z^i$ which leads to a simple constraint on the potential $G$.
Actually, the appropriate choice turns out to be $z^i = G^i$. Indeed it is straightfoward to show, using the 
Minkowski and stationarity conditions and the results (\ref{VIJ}), that
\be\label{eq}
V_{i \bar j} G^i G^{\bar j} = e^G \Big(6 - R_{i \bar j p \bar q}\, G^i G^{\bar j} G^p G^{\bar q} \Big)
\,.
\ee
This quantity must be positive if we want the matrix $V_{i\bar j}$ to be positive definite. 
Nevertheless it is important to stress that this is a necessary but not sufficient condition for stability.
Requiring (\ref{eq}) to be positive implies that:
\be
R_{i \bar j p \bar q}\, G^i G^{\bar j} G^p G^{\bar q} < 6\,.
\label{main}
\ee
Eq.~(\ref{main}) encodes our main results. Note that the curvature tensor $R_{i \bar j p \bar q}$ is determined 
by the second, third and fourth derivatives of $G$, but always mixing holomorphic and antiholomorphic indices 
so that it only depends on the K\"aler potential $K$. Therefore (\ref{main}) represents a bound 
on the values that the first derivatives $G_i$ (which depend on both $K$ and $W$) can take in terms of the curvature tensor. 
In addition to this bound we have also the constraint $G_{i \bar j} G^i G^{\bar j} = 3$ coming from the Minkowski condition. 
One can then imagine a situation with a fixed K\"ahler potential $K$ and an arbitrary superpotential $W$ 
(together with the constraint that the cosmological constant should vanish). This is equivalent to treat $G_{i \bar j}$ 
and $R_{i \bar j p \bar q}$ as fixed quantities and to scan over all the possible values of $G^i$ satisfying the restriction
$G_{i \bar j} G^i G^{\bar j} = 3$ and the bound (\ref{main}). It is then clear that eq.~(\ref{main}) puts constraints on the 
values that the ratios of the various $G^i$ can take, and actually requiring eq.~(\ref{main}) 
to have a solution also requires that $G_{i \bar j}$ and $R_{i \bar j p \bar q}$ satisfy 
certain conditions. This fact will become clear in the following sections. Indeed, the left-hand side of the inequality
(\ref{main}) is a function of the variables $G^i$, and since these variables 
take values over a compact set, as a consequence of (\ref{3}), this function has a finite 
minimum that depends only on $G_{i \bar j}$ and $R_{i \bar j p \bar q}$. An obvious necessary 
condition for the inequality (\ref{main}) to admit solutions is then that this minimum value should be smaller than 6. 
This implicitely defines a restriction that involves only $K$ and that is independent of the form of $W$. 
Unfortunately, since the inequality is a quartic polynomial in the $G^i$'s it seems difficult to derive the explicit form 
of such a condition in general\footnote{It is unlikely that this problem can be simplified by making a suitable choice
of K\"ahler frame and of holomorphic coordinate fields, along the lines of refs.~\cite{roceck},
as the crucial ingredient of (\ref{main}) is the curvature tensor. Note also that rewriting the
potential in terms of $\Omega = - 3\, e^{-G/3}$ and factorizing out a suitable positive definite
factor, along the lines of ref.~\cite{arthur}, one can obtain interesting alternative expressions
for the Minkowski, stationarity and stability conditions, that are equivalent to those used here.
However, this does not seem to simplify the task of deriving necessary conditions that depend
only on the geometry, like the one given by (\ref{main}).}. Explicit results can instead be easily derived in situations where 
the curvature has a rigidly fixed tensor structure and is controlled only by some scalar parameters. This is for instance
the case when the scalar manifold is the product of one-dimensional submanifolds associated to each field, or 
when the space is a symmetric space with sufficiently many isometries. We will concentrate on the first 
class of situations in the following sections.

\section{Separable K\"ahler potentials}
\setcounter{equation}{0}

As we already mentioned, there exists a mild assumption that can be made in order to simplify the study of the condition
(\ref{main}). It consists in assuming that the K\"ahler potential is separable into a sum of terms, each of them 
depending on a single field, while the superpotential can instead still be arbitrary:
\begin{eqnarray}
\label{simpk}
\begin{array}{lll}
K \!\!\!&=&\!\!\! \displaystyle{\sum_{k=1}^n K^{(k)}(\Phi_k,\Phi_k^\dagger)} \,, \smallskip\ \\ 
W \!\!\!&=&\!\!\! W(\Phi_1,\dots,\Phi_n) \,.
\end{array}
\end{eqnarray}
This assumption represents a K\"ahler-invariant constraint on the function $G$, implying that 
all its mixed derivatives vanish unless they are purely holomorphic or antiholomorphic. More concretely, 
it is straightforward to derive from (\ref{simpk}) that  
\begin{eqnarray}
\label{rest}
\begin{array}{lll}
\!\!\!&&\!\!\! G_{i \bar j}=0\,, \hspace{4cm} \; i,j \;\mbox{not equal}\,, \smallskip\ \\
\!\!\!&&\!\!\! G_{i j \bar k}=G_{i \bar j \bar k}=0\,, \hspace{2.5cm} \; i,j,k \;\mbox{not equal} \,, \smallskip\ \\
\!\!\!&&\!\!\! G_{i j k \bar l}=G_{i j \bar k \bar l}= G_{i \bar j \bar k \bar l} = 0\,,
\hspace{0.8cm} \;  \; i,j,k,l \;\mbox{not equal} \,, \smallskip \\
\!\!\!&&\!\!\! \dots
\end{array}
\end{eqnarray}
In particular, the K\"ahler metric computed from (\ref{simpk}) becomes diagonal. In fact, the whole 
K\"ahler manifold parametrized by the scalar fields factorizes into the product of $n$ K\"ahler 
submanifolds. The only non-vanishing components of the Riemann tensor are then the $n$ totally
diagonal components $R_{i\bar ii\bar i}$, which can furthermore be written as 
$R_{i\bar ii\bar i}=G_{i\bar i}^2 R_i$, where $R_i$ are the curvature scalars of the one-dimensional 
submanifolds associated to each of the fields:
\begin{equation}
\label{r}
R_i =  \frac {G_{i i \bar i \bar i}}{G_{i \bar i}^2} - \frac {G_{i i \bar i} G_{i i \bar i}}{G_{i \bar i}^3}
\,,\hspace{1cm}i=1,\cdots,n\,.
\end{equation}
It is clear that in this more restrictive situation the inequality (\ref{main}) simplifies substantially,
and that in this case its implications can be worked out in full generality, as we will show in the following 
sections.

Before starting our analysis, let us briefly recall the physical meaning of the quantities appearing in 
the theory. The overall supersymmetry breaking scale is parametrized by the gravitino mass 
$m_{3/2} = e^{G/2}$. The vacuum expectation values of the scalar fields belonging to each chiral 
superfield $\Phi_i$ are generically of order one, $\phi_i \sim 1$. The corresponding auxiliary 
fields are instead of the same order as the gravitino mass, $F_i \sim m_{3/2}$. Indeed, their values 
are given by $F_i = m_{3/2} {G_i}/{G_{i \bar i}}$, and the value of the scalar potential at the 
minimum can thus be written simply as $V = \sum_k G_{k \bar k}|F_k|^2 - 3 m_{3/2}^2$. In this way it 
becomes clear that the condition of vanishing cosmological constant forces all the auxiliary fields 
to be at most of order $m_{3/2}$. Finally, the Lagrangian for the fluctuations 
around the vacuum has the form:
\begin{equation}
{\cal L} = \sum_k G_{k \bar k}\, \partial_\mu \phi_k \partial^\mu \phi^{k*} 
- \sum_{k,l} \Big(V_{k \bar l}\, \phi^k \phi^{l*} + V_{k l}\, \phi^{k} \phi^{l} + {\rm c.c.}\Big) \,.
\end{equation}
The physical mass matrix for the scalar fields of the theory is thus obtained by rescaling the fields 
in such a way that their kinetic terms are canonically normalized. Doing that one finds:
\begin{equation}
M^2 = \left(
\begin{matrix}
m^2_{i \bar j} &  m^2_{ij} \smallskip \cr
m^2_{\bar i \bar j} & m^2_{\bar i j}
\end{matrix}
\right) \,,
\label{massmatrix}
\end{equation}
where the various entries are obtained by rescaling the second derivatives of the potential
with appropriate powers of the positive definite metric: 
\begin{eqnarray}
\begin{array}{lll}
\!\!\!&&\!\!\! \displaystyle{m^2_{i \bar j} = \frac {V_{i \bar j}}{\sqrt{G_{i \bar i} G_{j \bar j}}} \,,\;\;
m^2_{\bar i j} = \frac {V_{\bar i j}}{\sqrt{G_{i \bar i} G_{j \bar j}}}} \,,\;\; \smallskip\ \\
\!\!\!&&\!\!\! \displaystyle{ m^2_{i j} = \frac {V_{i j}}{\sqrt{G_{i \bar i} G_{j \bar j}}} \,,\;\;
m^2_{\bar i \bar j} = \frac {V_{\bar i \bar j}}{\sqrt{G_{i \bar i} G_{j \bar j}}}} \,.
\end{array}
\label{massmatrixcomp}
\end{eqnarray}

Let us now go on with our analysis of the necessary conditions for stability. In order to do so it is important 
to note that, under the assumption (\ref{simpk}) of separability of the K\"ahler potential, eqs.~(\ref{3}), (\ref{4}) 
and (\ref{vij}) substantially simplify. Actually the relevant formulas needed for our analysis can be written in 
a more transparent way making explicit their dependence on the function $G$. For later convenience, we 
will review the derivation done in the previous section for this particular case of theories. 

Our starting point is again the scalar potential (\ref{genpot}) but now for a function $G$ satisfying the properties 
(\ref{rest}). Using this the potential takes the form:
\begin{equation}
V=e^{G}\left(\sum_{k=1}^n \frac{G_{k} G_{\bar k}}{G_{k\bar{k}}}-3\right)\,.
\label{potential}
\end{equation}
Therefore the condition of vanishing cosmological constant $V=0$ implies in this particular case that the following 
constraint should hold on the vacuum:
\begin{equation}
\sum_{k=1}^n \frac{G_{k} G_{\bar k}}{G_{k\bar{k}}} = 3 \,.
\label{cosm}
\end{equation}
This represents one real condition among the first holomorphic derivatives $G_i$, which 
shall be interpreted as being the result of a tuning of the parameters of the theory. 
Its solution can be conveniently parametrized by introducing $n$ spherical parameters 
$\Theta_i$ subject to the constraint $\sum_k \Theta_k^2 = 1$ to describe the direction 
of the Goldstino in the $n$-dimensional field space, as well as $n$ arbitrary phases $\eta_i$
\cite{bimsoft}. One then simply finds that the first derivatives must have the form
\be
\label{cosmsol}
G_{i} = \sqrt{3}\, \eta_i\, \Theta_i\, \sqrt{G_{i \bar i}}
\ee
With this parametrization, the supersymmetry-breaking vacuum expectation values of the auxiliary fields 
of the chiral multiplets are given by $F_i = \sqrt{3}\,\eta_i\, \Theta_i\,G_{i \bar i}^{-1/2} m_{3/2}$.

The first derivatives of the potential with respect to the scalar fields are given, after using 
(\ref{cosm}), by the following expressions:
\begin{equation}
\label{derV}
V_i = e^G \left(G_i - \frac {G_i G_{\bar i} G_{i i \bar i}}{G_{i \bar i}^2} 
+ \sum_{k=1}^n \frac {G_{ik} G_{\bar k}}{G_{k \bar k}}\right) \,.
\end{equation}
Then the conditions of stationarity of the potential $V_i = 0$ imply that at the extremum:
\begin{equation}
\label{statprel}
\sum_{k=1}^n \frac {G_{ik} G_{\bar k}}{G_{k \bar k}} =
- G_i + \frac {G_i G_{\bar i} G_{i i \bar i}}{G_{i \bar i}^2} \,.
\end{equation}
These represent $n$ complex conditions on the second holomorphic derivatives, which in 
general fix the values of all the $n$ complex scalar fields. To proceed, let us assume for 
the moment that none of the $G_i$'s is identically zero. The conditions (\ref{statprel}) 
can then be used to express the second holomorphic derivatives $G_{ii}$ in terms of other 
types of derivatives:
\begin{equation}
G_{ii} = - \frac {G_{i \bar i}}{G_{\bar i}} 
\left(G_i + \sum_{k\neq i} \frac {G_{ik} G_{\bar k}}{G_{k \bar k}}\right) 
+ \frac {G_i G_{i i \bar i}}{G_{i \bar i}} \,.
\label{stat}
\end{equation}
The situation where some of the fields $\phi_\alpha$ (with $\alpha = 1,\dots,m$) preserve supersymmetry 
and lead to $G_\alpha=0$, while the rest of the fields $\phi_r$ (with $r=m+1,\dots,n$) break supersymmetry
and lead to $G_r\neq 0$, is slightly more subtle. It is however possible to study this special situation as a
particular subcase of the more general situation where all the fields break 
supersymmetry\footnote{Notice that, as we are looking for non-supersymmetric vacua satisfying (\ref{cosm}), 
we are excluding the case where all the $G_i$'s are equal to zero.}. Notice in this respect that such 
a case can exist only if, on top of 
the $n$ stationarity conditions (\ref{statprel}), also the $m$ additional constraints $G_\alpha=0$ are imposed, 
that is, if the theory satisfies $m$ additional restrictions (see for instance \cite{Ferrara:1994kg,Derendinger:2005ed} 
for examples of this type). More precisely, the stationarity conditions for the fields $\phi_\alpha$ are identically 
solved by setting $G_\alpha = 0$ only if the $m$ constraints $\sum_r G_{\alpha r} G_{\bar r}/G_{r \bar r} = 0$, 
which we shall think of as constraints on the superpotential, are fulfilled. Nevertheless, these situations 
can be obtained from the general case where none of the $G_i$'s vanishes identically by imposing the restrictions 
$\sum_r G_{\alpha r} G_{\bar r}/G_{r \bar r} = 0$ and taking then the limit $G_\alpha \to 0$. We will further comment 
on this in Sections 4 and 5.

The equations (\ref{cosmsol}) and (\ref{stat}) express the holomorphic first and second derivatives 
$G_i$ and $G_{ii}$ in terms of the Goldstino parameters $\Theta_i$ and $\eta_i$, the components 
of the metric $G_{i \bar i}$ and its derivatives $G_{i i \bar i}$, and the mixed holomorphic second 
derivatives $G_{ij}$ with $i\neq j$. They assure us that the point under consideration is an extremum 
of the potential with a vanishing cosmological constant.  
Now, to ensure local stability, we need to compute the Hessian matrix evaluated at the vacuum point, 
and check whether it is positive definite or not. The components of this matrix can be obtained by taking 
the second derivatives of (\ref{derV}) and simplifying the resulting expressions with the help of (\ref{cosmsol}) 
and (\ref{stat}). In order to write the results in a compact form, it is useful to introduce the notation 
$A_{ij} \equiv G_{ij}/G_{i} G_{j}$, $A_{i \bar j} \equiv G_{i \bar j}/G_{i} G_{\bar j}$, etc .... 
In this way one finds:
\begin{eqnarray}
\begin{array}{lll}
V_{i \bar i} \!\!\!&=&\!\!\! \displaystyle{G_i G_{\bar i}\,e^G 
\left[2 A_{i \bar i} - R_i + \sum_{k \neq i} \frac {|A_{ik}|^2}{A_{k \bar k}}
+ A_{i \bar i} \bigg(\sum_{k \neq i} \frac {A_{ik} \!+\! A_{\bar i \bar k}}{A_{k \bar k}}
+ \bigg|\sum_{k \neq i} \frac {A_{ik}}{A_{k \bar k}}\bigg|^2\bigg) \right]} \,,\hspace{-20pt} \smallskip\ \\
V_{i \bar j} \!\!\!&=&\!\!\! \displaystyle{G_i G_{\bar j}\,e^G
\left[- A_{i j} - A_{\bar i \bar j} - \sum_{k \neq i} \frac {A_{ik} A_{\bar i \bar j}}{A_{k \bar k}}
- \sum_{k \neq j} \frac {A_{\bar j \bar k} A_{i j}}{A_{k \bar k}}
+ \sum_{k \neq i,j} \frac {A_{ik} A_{\bar j \bar k}}{A_{k \bar k}}\right]} \,,\;\; 
i\neq j\,,\hspace{-20pt} \smallskip\ \\
V_{ii} \!\!\!&=&\!\!\! \displaystyle{G_i G_i\,e^G \left[- 2 A_{i \bar i} + 3 \frac {A_{i i \bar i}}{A_{i \bar i}} 
- \frac {A_{i i i \bar i}}{A_{i \bar i}^2} - 2 A_{i\bar i} \sum_{k \neq i} \frac {A_{ik}}{A_{k \bar k}} 
\bigg(1 - \frac {A_{i i \bar i}}{A_{i \bar i}^2} \bigg)
+ \sum_{k=1}^n \frac {A_{iik}}{A_{k \bar k}} \right]} \,, \hspace{-5pt}\smallskip\ \\
V_{i j} \!\!\!&=&\!\!\! \displaystyle{G_i G_j\,e^G \left[2 A_{ij} 
- \bigg(\frac {A_{i i \bar i}}{A_{i \bar i}^2} + \frac {A_{j j \bar j}}{A_{j \bar j}^2} \bigg) A_{ij}
+ \sum_{k=1}^n \frac {A_{ijk}}{A_{k \bar k}}\right]}\,, \hspace{3.2cm} i\neq j\,.\hspace{-10pt}
\end{array}
\label{hessiana}
\end{eqnarray}

The equations (\ref{hessiana}) represent, for theories satisfying the restriction (\ref{simpk}), the form 
of the general results (\ref{VIJ}) but showing the explicit dependence of the derivatives of the potential 
on the function $G$. Actually, as we already anticipated, the necessary condition (\ref{main}) obtained by imposing 
$V_{i \bar j} G^i G^{\bar j} > 0$ substantially simplifies in this case and can be written as 
\be
\sum_{k} R_k \bigg(\frac {G_k G_{\bar k}}{G_{k \bar k}} \bigg)^2 < 6 \,.
\label{mainsep}
\ee
This inequality is quadratic in the variables inside the brackets which are just three times the squared 
Goldstino angles and which also appear in the constraint (\ref{cosm}). This fact will allow us to derive in 
an exact way the constraints implied by (\ref{cosm}) and (\ref{mainsep}). To illustrate this, we will first 
consider the one-field and two-field cases, where exact information on the eigenvalues of the mass matrix 
can actually be obtained, and then we will consider the general $n$ field case\footnote{From now on we will 
consider the mass matrix instead of the Hessian matrix, as it is physically more relevant.}.

\section{Models with one field}
\setcounter{equation}{0}

Let us firstly consider the case of supergravity models involving a single chiral superfield $X$,
with arbitrary K\"ahler potential $K=K(X, X^\dagger)$ and arbitrary superpotential $W=W(X)$. 
This simple case might be also relevant to describe more complicated models with several 
superfields when for some reason one of the fields is much lighter than the others and therefore 
dominates supersymmetry breaking effects through its effective dynamics.

In the one-field case, we just have one frozen Goldstino parameter $\Theta_X=1$ and one arbitrary 
phase $\eta_X$. The Minkowski and stationarity conditions can be read from (\ref{cosmsol}) and (\ref{stat}),
and the supersymmetry-breaking auxiliary field is given by:
\begin{equation}
F_X = \sqrt{3}\,\eta_X\, G_{X \bar X}^{-1/2}\, m_{3/2} \,.
\end{equation}
The two independent components of the two-by-two mass matrix 
are found to be:
\begin{eqnarray}
\begin{array}{lll}
m^2_{X\bar X} \!\!\!&=&\!\!\! \Big(2-3\, R_X \Big) m_{3/2}^2 \,, \smallskip\ \\
m^2_{XX} \!\!\!&=&\!\!\! \eta_X^2\,\Big(\!-2+9 A_{XXX}+27 A_{XX\bar X}-27A_{XXX\bar X}\Big) m_{3/2}^2 \,.
\end{array}
\end{eqnarray}

In this case, the only quantity that depends on $W$ in the mass matrix is $A_{X X X}$.
The off-diagonal element $m^2_{X X}$ depends thus on both $K$ and $W$. The diagonal 
element $m^2_{X \bar X}$, on the other hand, depends only on $K$, and in fact only on the 
associated curvature $R_X$. The necessary condition (\ref{pos}) for local stability becomes just
$m^2_{X\bar X} > 0$, and (\ref{mainsep}) turns into a single very simple condition on $R_X$ of 
the form:
\begin{eqnarray}
R_X < \frac 23 \,.
\label{necess1}
\end{eqnarray}
This implies that $K$ should have curvature less than $2/3$, independently of $W$. 
In view of the form that the generalization of this result will take for several fields, 
it is however more appropriate to formulate it in terms of the inverse of the curvature. 
Assuming positive curvature, the condition takes the form:
\begin{eqnarray}
R_X^{-1} > \frac32 \,.
\label{constraintR1}
\end{eqnarray}

In this simplest case, the two eigenvalues of the mass matrix can actually be 
computed exactly. They are given by
\begin{equation}
\label{mass1}
m_\pm^2 = m^2_{X \bar X} \pm |m^2_{X X}| \,.
\end{equation}
This clearly shows that in order to be sure that the two eigenvalues are both positive, that is, 
in order to really have stability, one must switch from the simple necessary condition 
$m^2_{X \bar X}>0$, which involves only $K$, to the necessary and sufficient condition 
$m^2_{X \bar X} > |m^2_{X X}|$, which also involves $W$.

\section{Models with two fields}
\setcounter{equation}{0}

Let us consider next the slightly more complicated (but more representative) case of models 
involving two chiral superfields $X$ and $Y$, with separable K\"ahler potential 
$K = K^{(X)}(X,X^\dagger) + K^{(Y)}(Y,Y^\dagger)$ and arbitrary superpotential $W = W(X,Y)$.

In this case we need to introduce two constrained Goldstino parameters of the form $\Theta_X=\cos \theta$ 
and $\Theta_Y=\sin\theta$, and two arbitrary phases $\eta_X$ and $\eta_Y$. As in the previous case 
the Minkowski and stationarity conditions can be read from (\ref{cosmsol}) and (\ref{stat}).
Also, the two supersymmetry-breaking auxiliary fields are given by:
\begin{eqnarray}
\begin{array}{lll}
F_X \!\!\!&=&\!\!\! \displaystyle{\sqrt{3}\,\eta_X \cos \theta \, G_{X \bar X}^{-1/2}\, m_{3/2}} \,, \smallskip\ \\
F_Y \!\!\!&=&\!\!\! \displaystyle{\sqrt{3}\,\eta_Y\, \sin \theta \, G_{Y\, \bar Y\,}^{-1/2}\, m_{3/2}} \,.
\end{array}
\end{eqnarray}
The six independent components of the canonically normalized four-by-four mass matrix 
are found to be:
\begin{eqnarray}\label{eq2}
\begin{array}{lll}
m^2_{X \bar X} \!\!\!&=&\!\!\! \bigg[2 - 3\,\cos^2 \theta\, R_X 
+ 3\,\sin^2 \theta\, \Big(A_{X Y} + A_{\bar X \bar Y} + 3\, \big|A_{X Y}\big|^2\Big)\bigg]\, m_{3/2}^2 \,, \smallskip\ \\
m^2_{Y\, \bar Y\,} \!\!\!&=&\!\!\! \bigg[2 - 3\,\sin^2 \theta\, R_Y 
+ 3\,\cos^2 \theta\, \Big(A_{X Y} + A_{\bar X \bar Y} + 3\, \big|A_{X Y}\big|^2\Big)\bigg]\, m_{3/2}^2 \,, \smallskip\ \\
m^2_{X \bar Y\,} \!\!\!&=&\!\!\!  \displaystyle{\frac {\eta_X}{\eta_Y} \bigg[\!\!-\!3\,\sin \theta\cos \theta\, 
\Big(A_{X Y} + A_{\bar X \bar Y} + 3\, \big|A_{X Y}\big|^2\Big)\bigg]\, m_{3/2}^2} \,, \smallskip\ \\
m^2_{X X} \!\!\!&=&\!\!\! \eta_X^2 \bigg[\!\!-\!2 + 9 \cos^4 \theta\, A_{X X X} + 27 \cos^4 \theta\, A_{X X \bar X} 
- 27 \cos^6 \theta\, A_{X X X \bar X} \smallskip\ \\
\!\!\!&\;&\!\!\! \hspace{18pt} -\, 6 \sin^2 \theta \, \Big(1 - 9 \cos^4 \theta\, A_{X X \bar X}\Big) A_{X Y\,}\!
+ 9 \sin^2 \theta \cos^2 \theta\, A_{X X Y\,}\!\bigg]\, m_{3/2}^2 \,, \smallskip\ \\
m^2_{Y\, Y\,} \!\!\!&=&\!\!\! \eta_Y^2 \bigg[\!\!-\!2 + 9 \sin^4 \theta\, A_{Y\, Y\, Y\,} + 27 \sin^4 \theta\, A_{Y\, Y\, \bar Y\,}\!
- 27 \sin^6 \theta\, A_{Y\, Y\, Y\, \bar Y\,}  \smallskip\ \\
\!\!\!&\;&\!\!\! \hspace{18pt} -\, 6 \cos^2 \theta \, \Big(1 - 9 \sin^4 \theta\, A_{Y\, Y\, \bar Y\,}\!\Big) A_{X Y\,}\!
+ 9 \sin^2 \theta \cos^2 \theta\, A_{Y\, Y\, X}\bigg]\, m_{3/2}^2 \,, \smallskip\ \\
m^2_{X Y\,} \!\!\!&=&\!\!\! \eta_X \eta_Y \bigg[ 3\, \sin \theta\cos \theta\, 
\Big(2 - 9 \cos^4\theta\,A_{X X \bar X} - 9 \sin^4\theta\,A_{Y\, Y\, \bar Y\,}\!\Big) A_{X Y\,}\! \smallskip\ \\
\!\!\!&\;&\!\!\! \hspace{32pt} +\, 9 \cos^3\theta\sin\theta\, A_{X X Y\,} + 9 \sin^3\theta\cos\theta\, A_{Y\, Y\, X}
\bigg]\, m_{3/2}^2 \,.
\end{array}
\end{eqnarray}

In this case there are several quantities that dependent on $W$ in the mass matrix: $A_{X X X}$,
$A_{Y\, Y\, Y\,}$, $A_{X X Y\,}$, $A_{Y\, Y\, X}$ and also $A_{X Y\,}$. As can be seen from (\ref{eq2}) the 
``off-diagonal elements" $m^2_{X X}$, $m^2_{Y\, Y\,}$ and $m^2_{X Y\,}$ depend heavily on both $K$ and $W$ 
but the ``diagonal elements" $m^2_{X \bar X}$, $m^2_{Y\, \bar Y\,}$ and $m^2_{X \bar Y\,}$, on the other hand, 
depend mostly on $K$, with only a mild dependence on $W$ arising through terms involving just $A_{X Y\,}$. 
The necessary conditions (\ref{main}) for positive definiteness for the Hessian matrix imply that the 
mass matrix should fulfill the condition $\sum_{i,j} \eta_i \Theta_i \eta_j^* \Theta_j m_{i \bar j}^2> 0$. 
Actually in this two-fields case it is easy to see using (\ref{eq2}) how this leads to a 
simple condition where the dependence on $A_{X Y}$ in the various components of $m_{i \bar j}^2$  
cancels out. The condition finally takes the form (\ref{mainsep}) and reads:
\be
\label{i1}
\cos^4 \theta\,R_X  + \sin^4 \theta\, R_Y < \frac23 \,.
\ee

Assuming for simplicity that the curvatures are positive, it is straightforward to verify that the 
inequality (\ref{i1}) has solutions only if the constraint
\begin{equation}
R_X^{-1} + R_Y^{-1} > \frac32
\label{constraintR2}
\end{equation}
is satisfied, and the angle $\theta$ is restricted to be within the interval:
\begin{equation}
\theta \in [\theta_{\rm min},\theta_{\rm max}] \,,
\label{constrainttheta2}
\end{equation}
where
\begin{eqnarray}
\begin{array}{lll}
\theta_{\rm min} \!\!\!&=&\!\!\!
\left\{\begin{array}{lll}
\displaystyle{{\rm arccos}\, 
\sqrt{\frac {R_X^{-1} \!+\! \sqrt{R_X^{-1} R_Y^{-1}(R_X^{-1} \!+\!R_Y^{-1}\!-\! 3/2)/(3/2)}}{R_X^{-1} \!+\! R_Y^{-1}}}} \,,
&\mbox{if}& R_X^{-1} < 3/2\,, \hspace{-30pt} \bigskip\ \\[5mm]
0\,, &\mbox{if}& R_X^{-1} > 3/2\,.  \hspace{-30pt} \smallskip\ \\
\end{array}\right. \bigskip\ \\
\theta_{\rm max} \!\!\!&=&\!\!\!
\left\{\begin{array}{lll}
\displaystyle{{\rm arcsin}\, 
\sqrt{\frac {R_Y^{-1} \!+\! \sqrt{R_X^{-1} R_Y^{-1}(R_X^{-1}\!+\!R_Y^{-1} \!-\! 3/2)/(3/2)}}{R_X^{-1} \!+\! R_Y^{-1}}}} \,,
&\mbox{if}& R_Y^{-1} < 3/2\,,  \hspace{-30pt} \bigskip\ \\
\displaystyle{\frac \pi 2}\,, &\mbox{if}& R_Y^{-1} > 3/2\,.  \hspace{-30pt}\smallskip\ \\
\end{array}\right. \smallskip\ \\
\end{array}
\label{thetacrit}
\end{eqnarray}
Notice that, given (\ref{constraintR2}), $\theta_{\rm min}$ and $\theta_{\rm max}$ are always real and they satisfy 
$\theta_{\rm min} < \theta_{\rm max}$. Also note that the constraint (\ref{constraintR2}) is clearly the generalization 
of the condition (\ref{constraintR1}) arising in the single field case.

Before going on with the analysis, it is important to point out the fact that the restriction 
(\ref{constrainttheta2}) on 
the angle is qualitatively different depending on the values of the inverse curvatures $R_X^{-1}$ and $R_Y^{-1}$ 
(provided they fulfilled the condition (\ref{constraintR2})). If $R_X^{-1} > 3/2$ and $R_Y^{-1} > 3/2$, then $\theta_{\rm min}=0$ 
and $\theta_{\rm max}=\pi/2$, and all the angles are allowed. If $R_X^{-1} > 3/2$ and 
$R_Y^{-1} < 3/2$, then $\theta_{\rm min}=0$ and $\theta_{\rm max}<\pi/2$, and
only angles that are smaller than a critical upper bound are allowed. If $R_X^{-1} < 3/2$ and $R_Y^{-1} > 3/2$, then 
$\theta_{\rm min}>0$ and $\theta_{\rm max}=\pi/2$, and only angles that are 
larger than a critical lower bound are allowed. Finally, if $R_X^{-1} < 3/2$ and $R_Y^{-1} < 3/2$, then $\theta_{\rm min}>0$ 
and $\theta_{\rm max} <\pi/2$, and only angles that are within some critical 
upper and lower bounds are allowed.

These results reflect the fact that there is in general an obstruction against achieving values of $\theta$ close 
to $0$ or $\pi/2$, corresponding to one of the two fields being the Goldstino, if that field 
does not satisfy on its own the necessary condition $R^{-1} > 3/2$, relevant for the single field case. This also implies 
that for given curvatures $R_X$ and $R_Y$, the ratio of the two supersymmetry-breaking auxiliary fields is constrained 
to lie in a certain region, since $|\sqrt{G_{Y \bar Y}} F_Y|/|\sqrt{G_{X \bar X}}F_X| = \tan \theta$.

In the general case of two fields with a complex mass matrix, the four exact eigenvalues of the mass 
matrix cannot be computed exactly. However, this can be done in the special case where all the entries are real. 
In this case one finds:
\begin{eqnarray}
\begin{array}{lll}
m^2_{1\pm} \!\!\!&=&\!\!\! 
\displaystyle{\frac 12 \Big(m^2_{X \bar X} + m^2_{X X} + m^2_{Y\, \bar Y\,} + m^2_{Y\, Y\,}\Big)} \smallskip\ \\
\!\!\!&\;&\!\!\! \displaystyle{\pm\, \frac 12 \sqrt{\Big(m^2_{X \bar X} + m^2_{X X} 
- m^2_{Y\, \bar Y\,} - m^2_{Y\, Y\,}\Big) ^2 + 4 \Big(m^2_{X \bar Y} + m^2_{X Y\,}\Big)^2}} \,, \smallskip\ \\
m^2_{2\pm} \!\!\!&=&\!\!\! 
\displaystyle{\frac 12 \Big(m^2_{X \bar X} - m^2_{X X} + m^2_{Y\, \bar Y\,} - m^2_{Y\, Y\,}\Big)} \smallskip\ \\
\!\!\!&\;&\!\!\! \displaystyle{\pm\, \frac 12 \sqrt{\Big(m^2_{X \bar X} - m^2_{X X} 
- m^2_{Y\, \bar Y\,} + m^2_{Y\, Y\,}\Big) ^2 + 4 \Big(m^2_{X \bar Y} - m^2_{X Y\,}\Big)^2}} \,.
\end{array}
\label{mass2}
\end{eqnarray}
As before, these equations clarify the fact that in order to be sure that all the four eigenvalues are really positive, 
one must switch from the simple necessary conditions derived above to much stronger and complicated necessary
and sufficient conditions.

So far in this section we have assumed that $G_X \neq 0$ and $G_Y \neq 0$, so that none of the 
auxiliary fields vanishes. It is however interesting and instructive to explore what can happen in the limit in which 
one of the two fields has a vanishing auxiliary field, that is, when $\theta$ approaches $0$ or $\pi/2$. 
For concreteness let us study the case in which $\theta \to 0$ (the case $\theta \to \pi/2$ is clearly analogous, 
with the roles of $X$ and $Y$ interchanged). At leading order in $\theta$ the auxiliary fields take the form
\begin{eqnarray}
\begin{array}{lll}
F_X \!\!\!&\simeq&\!\!\! \sqrt{3}\,\eta_X G_{X \bar X}^{-1/2}\, m_{3/2} \,, \smallskip\ \\
F_Y \!\!\!&\simeq&\!\!\! 0 \,.
\end{array}
\end{eqnarray}
To derive the leading behavior of the mass matrix notice that, according to their 
definitions, $A_{X X X}$, $A_{X X \bar X}$ and $A_{X X X \bar X}$ behave like $\theta^{0}$, 
$A_{X Y\,}$ and $A_{X X Y\,}$ like $\theta^{-1}$, $A_{Y\, Y\, X}$ like $\theta^{-2}$, 
$A_{Y\, Y\, Y\,}$ and $A_{Y\, Y\, \bar Y\,}$ like $\theta^{-3}$, and 
$A_{Y\, Y\, Y\, \bar Y\,}$ like $\theta^{-4}$. Keeping both the finite terms and the leading divergent 
terms, one gets:
\begin{eqnarray}
\begin{array}{lll}
m^2_{X \bar X} \!\!\!&\simeq&\!\!\! 
\displaystyle{\bigg[2 - 3\, R_X + \frac {|G_{X Y\,}|^2}{G_{X \bar X} G_{Y\, \bar Y\,}} \bigg]\, m_{3/2}^2} \,, 
\smallskip\ \\
m^2_{Y\, \bar Y\,} \!\!\!&\simeq&\!\!\! \displaystyle{\bigg[2 
+ \theta^{-2} \frac {|G_{X Y\,}|^2}{G_{X \bar X} G_{Y\, \bar Y\,}}\bigg]\, m_{3/2}^2} \,,
\smallskip\ \\
m^2_{X \bar Y\,} \!\!\!&\simeq&\!\!\!  \displaystyle{\frac {\eta_X}{\eta_Y} \bigg[
\!- \theta^{-1} \frac {|G_{X Y\,}|^2}{G_{X \bar X} G_{Y\, \bar Y\,}}\bigg]\, m_{3/2}^2} \,, 
\smallskip\ \\
m^2_{X X} \!\!\!&\simeq&\!\!\! \eta_X^2 
\bigg[\!-2 + 9\, A_{X X X} + 27\, A_{X X \bar X} - 27\, A_{X X X \bar X} \bigg]\, m_{3/2}^2 \,,
\smallskip\ \\
m^2_{Y\, Y\,} \!\!\!&\simeq&\!\!\! \displaystyle{\eta_Y^2
\bigg[\!-2 + 2\,\sqrt{3}\, \frac {G_{X Y\,}G_{Y\,Y\,Y\,}}{\sqrt{G_{X \bar X}} G_{Y \,\bar Y\,}^2} 
+ 3\, \frac {G_{Y\,Y\,X}}{\sqrt{G_{X \bar X}} G_{Y \,\bar Y\,}} \bigg]\, m_{3/2}^2} \,,
\smallskip\ \\
m^2_{X Y\,} \!\!\!&\simeq&\!\!\! \displaystyle{\eta_X \eta_Y
\bigg[3\,\sqrt{3}\,\frac {G_{X X Y\,}}{G_{X \bar X} \sqrt{G_{Y \,\bar Y\,}}} \bigg]\, m_{3/2}^2} \,.
\end{array}
\label{masseslimit}
\end{eqnarray}
At this point, there are two distinct situations (as discussed in Section 3) that can be considered, 
depending on whether or not the quantity $G_{X Y\,}$ vanishes or not. 

If $G_{X Y\,} \neq 0$, then sending $G_Y \to 0$ does not help in solving the stationarity conditions
and what happens is that the two scalar degrees of freedom in $Y$ become infinitely heavy and 
decouple, leaving only the field $X$ in the low-energy effective theory. More precisely, it is 
straightforward to verify that the four eigenvalues reduce in this limit to the following expressions:
\begin{eqnarray}
\begin{array}{lll}
m^2_{a+} \!\!\!&\simeq&\!\!\! \displaystyle{\bigg(m^2_{X \bar X} - \frac {|m^2_{X \bar Y}|^2}{m^2_{Y\, \bar Y\,}} \bigg) 
\pm |m^2_{X X}|}\,, \smallskip\ \\
m^2_{a-} \!\!\!&\simeq&\!\!\! m^2_{Y\, \bar Y\,} \,.
\end{array}
\end{eqnarray}
The first pair of eigenvalues are associated to the field $X$ (the one with sizeable auxiliary field) and are finite, 
as the factor $|m^2_{X \bar Y}|^2/m^2_{Y\, \bar Y\,}$ just cancels the last term in the factor $m^2_{X \bar X}$. 
This means that all the dependence on $G_{X Y\,}$ disappears and the same result as for the one field case 
is recovered. Note also that by taking the limit $\theta \to 0$ of eq.~(\ref{i1}), one directly recovers the necessary 
condition (\ref{constraintR1}) for the field $X$. The second pair of eigenvalues are instead associated to the field $Y$ 
(the one with vanishing auxiliary field) and diverge like $\theta^{-2}$, with a coefficient that is proportional to 
$|G_{X Y\,}|^2$ and always positive.

On the other hand, if $G_{X Y\,} = 0$, sending $G_Y \to 0$ does solve the stationarity conditions, and the masses of the 
four degrees of freedom in $X$ and $Y$ are expected to all remain finite. This is manifestly true since in this case all the 
divergent terms disappear from (\ref{masseslimit}). There is however an important new feature 
that appears in this special situation: one finds $m^2_{X \bar Y\,} = 0$ and $m^2_{Y \bar Y\,} = 2$. 
The necessary condition (\ref{pos}) then collapses to the condition $m^2_{X \bar X} > 0$, which coincides again
with the necessary condition (\ref{constraintR1}) for the field $X$. None of the masses is however
automatically positive in this case. For instance, if one also has $G_{X X Y\,} = 0$ and $G_{Y\, Y\, X} = 0$, as is 
for example the case when the superpotential factorizes as $W(X,Y) = W^{(X)}(X) W^{(Y)}(Y)$, the four 
eigenvalues reduce to the following simple expressions:
\begin{eqnarray}
\label{eigenlimit}
\begin{array}{lll}
m^2_{a+} \!\!\!&\simeq&\!\!\! \displaystyle{m^2_{X \bar X} \pm |m^2_{X X}|}\,, \smallskip\ \\
m^2_{a-} \!\!\!&\simeq&\!\!\! \Big(2 \pm 2 \Big) m_{3/2}^2 \,.
\end{array}
\end{eqnarray}

This analysis shows that when one of the two complex fields has a negligible supersymmetry-breaking auxiliary 
field, then the necessary condition for local stability always collapses to the one obtained in the one-field case
for the field that breaks supersymmetry. The detailed form of the eigenvalues depends however
on whether $G_{X Y\,}$ vanishes or not. If $G_{X Y} \neq 0$, the field with negligible auxiliary field decouples and 
the whole problem reduces to a one-field problem. On the other hand, if $G_{X Y} = 0$ this field does not decouple 
and might lead to instabilities. 

Notice finally that the results obtained in this section for the two-field case imply that if a theory with a single field 
is described by a K\"ahler potential that does not satisfy the necessary condition (\ref{constraintR1}) for stability, 
it is still possible to achieve a stable situation by adding an additional field with a K\"ahler potential such that the 
two-field case necessary condition (\ref{constraintR2}) is satisfied. However, the extra field cannot be much heavier 
than the original one. Indeed, if this were the case one could integrate out the heavy field, getting only a small correction 
to the K\"ahler potential of the light field, which would be in principle not enough help for the light field to fulfill the necessary
condition (\ref{constraintR1}).

\section{Models with several fields}
\setcounter{equation}{0}

Let us finally consider the more general case of supergravity models that involve an arbitrary number $n$ of chiral
superfields $\Phi_i$, with a separable K\"ahler potential of the form $K=\sum_k K^{(k)}(\Phi_k, \Phi_k^\dagger)$
and an arbitrary superpotential $W=W(\Phi_1,\dots,\Phi_n)$. We will show that the constraints imposed
by the necessary condition (\ref{pos}) and (\ref{main}) for local stability are of the same type as those found in
the previous two sections. In particular, the constraint on the curvatures that was found in the one-field and
two-field cases,  see eqs.~(\ref{constraintR1}) and (\ref{constraintR2}), turns out to generalize in the expected
way to the $n$-field case. Similarly, the variables $\Theta_i$ parametrizing the Goldstino direction are 
constrained to a finite range of values depending on the curvatures, as in the two-field case
eqs.~(\ref{constrainttheta2}) and (\ref{thetacrit}).

As we already mentioned in Section 3, the Minkowski condition (\ref{cosmsol}) reflecting the vanishing of the 
cosmological constant can be solved in this general case by introducing $n$ angular variables $\Theta_i$ 
satisfying the constraint $\sum_k \Theta_k^2 = 1$ and $n$ arbitrary phases $\eta_i$.
The supersymmetry-breaking auxiliary fields are given by:
\begin{equation}
F_i = \sqrt{3}\,\eta_i\,\Theta_i\,G_{i \bar i}^{-1/2} m_{3/2} \,.
\end{equation}
The general results (\ref{hessiana}) for the components of the Hessian matrix can be written
in terms of the variables $\Theta_i$ and the phases $\eta_i$, and they lead to 
the following expressions for the squared masses:
\begin{eqnarray}
\begin{array}{lll}
m^2_{i \bar i} \!\!\!&=&\!\!\! \displaystyle{
\Bigg[2 - 3\, \Theta_i^2 R_i
+ 3 \sum_{k \neq i} \Theta_k^2 \Big(A_{ik} + A_{\bar i \bar k} +
3\,\Theta_i^2 \big|A_{ik}\big|^2 \Big)
+ 9 \bigg|\sum_{k \neq i} \Theta_k^2 A_{i k} \bigg|^2 \Bigg]
m_{3/2}^2} \,,\hspace{-20pt} \smallskip\ \\
m^2_{i \bar j} \!\!\!&=&\!\!\! \displaystyle{\frac {\eta_i}{\eta_j}
\Bigg[\!- 3\, \Theta_i \Theta_j \bigg(A_{i j} + A_{\bar i \bar j} + 3
\Big(\Theta_i^2 + \Theta_j^2 \Big) \big|A_{ij}\big|^2} \smallskip\ \\
\!\!\!&\;&\!\!\! \hspace{20pt} \displaystyle{+\, 3 \sum_{k \neq i,j}
\Theta_k^2 \Big(A_{ik} A_{\bar i \bar j} + A_{\bar j \bar k} A_{i j}
- A_{ik} A_{\bar j \bar k} \Big) \bigg)\Bigg] m_{3/2}^2} \,, 
\hspace{2.6cm} i\neq j \,,\hspace{-25pt} \smallskip\ \\
m^2_{ii} \!\!\!&=&\!\!\! \displaystyle{\eta_i^2 \Bigg[\!- 2 + 9\,
\Theta_i^4 A_{iii}
+ 27\, \Theta_i^4 A_{i i \bar i} - 27\, \Theta_i^6 A_{i i i \bar i}}
\smallskip\ \\
\!\!\!&\;&\!\!\! \hspace{19pt} \displaystyle{+\, 3\, \sum_{k \neq i}
\Theta_k^2
\bigg(\!\!-\!2\, A_{ik} \Big(1 - 9\, \Theta_i^4 A_{i i \bar i} \Big) +
3\,\Theta_i^2 A_{iik} \bigg) \Bigg]m_{3/2}^2} \,, \smallskip\ \\
m^2_{i j} \!\!\!&=&\!\!\! \displaystyle{\eta_i \eta_j \Bigg[3\,
\Theta_i \Theta_j \bigg( \Big(2 - 9\, \Theta_i^4 A_{i i \bar i}
- 9\, \Theta_j^4 A_{j j \bar j} \Big) A_{ij}
+ \sum_{k=1}^n \Theta_k^2 A_{ijk} \bigg)\Bigg] m_{3/2}^2}\,,
\hspace{0.3cm} i\neq j\,.\hspace{-25pt}
\end{array}
\label{matrixn}
\end{eqnarray}

The quantities that dependent on $W$ in the mass matrix are $A_{i j k}$ and $A_{i j}$. As in the previous cases 
note that the ``off-diagonal elements" $m^2_{i j}$ depend heavily on both $K$ and $W$, whereas the 
``diagonal elements" $m^2_{i \bar j}$ depend mostly on $K$, the only dependence on $W$ arising 
through terms involving just $A_{i j}$. As we already mentioned in the previous section, the necessary condition 
(\ref{main}) for positive definiteness of the mass matrix reads $\sum_{i,j} \eta_i \Theta_i \eta_j^* \Theta_j m_{i \bar j}^2> 0$. 
This leads again to a simple condition where all the dependence on the $A_{i j}$'s cancels out, which
corresponds to the inequality (\ref{mainsep}). This inequality can be rewritten in terms of the parameters $\Theta_i$ 
and takes the form:
\be
\sum_{i=1}^n \Theta_i^4 R_i < \frac 23 \,.
\label{constraintRThetan}
\ee
This expression generalizes the conditions (\ref{necess1}) and (\ref{i1}) that were obtained in 
one-field and two-field cases.

The constraint (\ref{constraintRThetan}) can be interpreted as an upper bound on the function $f(x_i) = \sum_i R_i\, x_i^2$, 
where the curvatures $R_i$ are treated as constants and the real variables $x_i=\Theta_i^2$ range from $0$ to $1$ and are 
subject to the constraint $\sum_k x_k = 1$. In particular, the inequality (\ref{constraintRThetan}) implies that $f_{\rm min}<2/3$, 
where $f_{\rm min}$ is the minimum value of $f(x_i)$ within the allowed range for the $x_i$. Finding $f_{\rm min}$ 
is a constrained minimization problem which can be solved in the standard way using Lagrangian multipliers. 
Assuming again for simplicity that all the curvatures $R_i$ are positive, it is straightforward to show that the values of the 
variables at the minimum are given by $x_i = R_i^{-1}/(\sum_k R_k^{-1})$, and therefore $f_{\rm min} = 1/(\sum_k R_k^{-1})$. 
The condition $f_{\rm min} < 2/3$ then implies that the constraint on the curvatures takes the form:
\be
\sum_{k=1}^n R_k^{-1} >\frac 32 \,.
\label{constraintRn}
\ee
If the curvatures satisfy the restriction (\ref{constraintRn}), then the condition (\ref{constraintRThetan}) 
admits solutions, but only for a limited range of values for the variables $\Theta_i$. These ranges can 
be easily determined by proceeding as follows: We first rewrite (\ref{constraintRThetan}) in the form
$f(x_i) < 2/3$ and use the constraint $\sum_k x_k = 1$ to eliminate one of the variables and work with
$n-1$ unconstrained variables. As the function $f$ is a concave parabola with respect to any of the variables,
a given variable is allowed to vary in a range that is bounded by the two solutions of the quadratic equation 
$f = 2/3$. In order for this interval to be non-empty, however, these solutions must be real, meaning that the 
argument of the square-root appearing in them must be positive. This represents a new inequality similar to 
the one we started with, but with one less variable. Repeating then iteratively the same procedure, one
eventually arrives to an inequality involving a single variable. The two solutions of the corresponding equality 
define then the range that this variable is allowed to take in terms of the curvatures $R_i$, and it is real 
provided that the condition (\ref{constraintRn}) holds. The final result is that the spherical variables 
$\Theta_i$ parametrizing the Goldstino direction are constrained 
as 
\be
\Theta_i \in [\Theta_{i-},\Theta_{i+}]
\label{constraintthetan}
\ee
where
\begin{eqnarray}
\begin{array}{lll}
\Theta_{i+} \!\!\!&=&\!\!\!
\left\{\begin{array}{lll}
\displaystyle{
\sqrt{\frac {R_i^{-1} \!+\! \sqrt{R_i^{-1} (\sum_{k \neq i} \! R_k^{-1})
(\sum_k \! R_k^{-1} \!-\! 3/2)/(3/2)}}{\sum_k \! R_k^{-1}}}} \,,
&\mbox{if}& R_i^{-1} < 3/2\,, \hspace{-30pt} \bigskip\ \\
1\,, &\mbox{if}& R_i^{-1} > 3/2\,. \hspace{-30pt} \smallskip\ \\
\end{array}\right. \bigskip\ \\
\Theta_{i-} \!\!\!&=&\!\!\!
\left\{\begin{array}{lll}
\displaystyle{
\sqrt{\frac {R_i^{-1} \!-\! \sqrt{R_i^{-1} (\sum_{k \neq i} \! R_k^{-1})
(\sum_k \! R_k^{-1} \!-\! 3/2)/(3/2)}}{\sum_k \! R_k^{-1}}}} \,,
&\mbox{if}& \sum_{k \neq i} \! R_k^{-1} < 3/2\,, \hspace{-30pt} \bigskip\ \\
0\,, &\mbox{if}& \sum_{k \neq i} \! R_k^{-1} > 3/2\,. \hspace{-30pt} \smallskip\ \\
\end{array}\right. \smallskip\ \\
\end{array}
\label{thetacritn}
\end{eqnarray}
These expressions generalize eqs.~(\ref{constrainttheta2}) and (\ref{thetacrit}) found in the 
two-field case, for which $\Theta_1 = \cos \theta$, $\Theta_2 = \sin \theta$, and therefore 
$\Theta_{1-} = \cos \theta_{\rm max}$, $\Theta_{1+} = \cos \theta_{\rm min}$ and 
$\Theta_{2-} = \sin \theta_{\rm min}$, $\Theta_{2+} = \sin \theta_{\rm max}$. Again, they
imply that the ratios of the auxiliary fields of the various fields are constrained as well,
since $|\sqrt{G_{i \bar i}} F_i|/|\sqrt{G_{j \bar j}} F_j| = \Theta_i/\Theta_j$.

Although the analysis done so far in this section for the $n$-field case formally excludes the 
particular situations of the type mentioned in Section 3, where some of the fields could have 
an identically vanishing auxiliary field, it is possible to extract information about these cases
by taking careful limits of the more general case. More concretely, let us consider the limit in 
which $m$ of the $\Theta_i$'s are sent to zero, and the remaining $n-m$ are kept finite: 
$\Theta_\alpha \to 0$, $\Theta_r \neq 0$, $\sum_r \Theta_r^2 \to 1$. 
As already explained in Section 3, this implies that $G_\alpha = 0$, which however solves 
the stationarity condition with respect to the field $\phi_\alpha$ only if the extra condition
$\sum_r G_{\alpha r} G_{\bar r}/G_{r \bar r} = 0$ is satisfied. If this constraint is not satisfied 
by the theory, then the corresponding field must decouple in the limit that is considered, and 
the net outcome is a reduction of the number of relevant fields. If instead it is satisfied, the 
corresponding field $\phi_\alpha$ can remain light, and must be kept in the analysis. 
Nevertheless, the inequality (\ref{constraintRThetan}) that is at the origin of our necessary 
conditions simplifies, as all the fields that have a negligible auxiliary field drop out. One 
therefore obtains exactly the same constraint on the curvatures and the Goldstino direction 
as before, but just restricted to those fields that have a significant auxiliary field. In any case, 
one therefore concludes that the fields that have negligibly small auxiliary fields do not 
influence the necessary conditions for local stability that we derived.

\section{Moduli fields in string models}
\setcounter{equation}{0}

String models provide supergravity low-energy effective theories that certainly count among the
most promising and motivated candidates for supersymmetric extensions of the standard model.
Actually, the moduli fields arising in string compactifications to four dimensions seem to be 
natural candidates to constitute the hidden sector of the theory that is  supposed to be responsible 
for supersymmetry breaking. The K\"ahler potential and superpotential governing the dynamics of 
these moduli fields typically have the general structure
\begin{eqnarray}
\label{stringk}
\begin{array}{lll}
K \!\!\!&=&\!\!\! \displaystyle{- \sum_{a=1}^n n_a\, {\rm ln} (\Phi_a + \Phi_a^\dagger) + \dots} \,, \smallskip\ \\
W \!\!\!&=&\!\!\! W(\Phi_1,\dots,\Phi_n) \,,
\end{array}
\end{eqnarray}
where by the dots we denote corrections that are subleading in the derivative and loop 
expansions defining the effective theory. The moduli sector therefore fulfills the assumption 
(\ref{simpk}), and the 
general necessary condition (\ref{constraintRn}) for the local stability of any non-supersymmetric 
Minkowski vacuum applies\footnote{It should be noted that charged fields usually have strong mixings 
with moduli in the K\"ahler potential. However, we shall assume here that these fields do not participate to 
supersymmetry breaking, and can therefore be ignored, on the same footing as matter fields.}. 
The K\"ahler curvatures can be computed using (\ref{r}) and in this 
particular case they take constant values given just by $R_i = 2/n_i$. 
The necessary condition (\ref{constraintRn}) thus implies the very simple and strong restriction:
\be
\sum_{k=1}^n n_k > 3 \,.
\label{necessstring}
\ee
In the simplest case involving just a single modulus, this results was already derived in 
ref.~\cite{Brustein:2004xn}, although in a less direct way.

The result (\ref{necessstring}) puts severe restrictions on the situations where a single modulus dominates
the dynamics. For instance, the universal dilaton $S$ has $n_S = 1$ and therefore does not fulfill the necessary 
condition (\ref{necessstring}). Subleading corrections to the K\"ahler potential cannot help in this case, since if 
they are to be small 
they cannot radically modify the K\"ahler curvature. We therefore conclude that the scenario proposed in 
ref.~\cite{dilatondom}, in which the dilaton dominates supersymmetry breaking, can never be realized in a 
controllable way\footnote{Similar conclusions were already reached in the past, but relying on the additional 
assumption that the vacuum expectation value of $S$ should be large in order to have weak string coupling, 
and using different arguments. For instance, it was argued in ref.~\cite{Casas:1996zi} that even if local 
Minkowski minima can arise, they cannot be global minima, and in ref.~\cite{Brustein:2000mq} it was 
shown that local Minkowski minima cannot be realized if the superpotential is assumed to be steep. 
See also ref.~\cite{Burgess:1995aa, Barreiro:1997rp, bgw} for other relevant discussions concerning 
non-perturbative corrections to the dilaton K\"ahler potential}. 
On the other hand, the overall K\"ahler modulus $T$ has $n_T = 3$, and violates 
only marginally the necessary condition when considered on its own. In this case, subleading corrections to 
the K\"ahler potential are crucial, since even a slight change in the curvature can allow this field to fulfill the 
necessary condition. Recently, the form of this corrections has been better investigated in various classes of string models 
and some interesting cases where they can help achieving a satisfactory scenario based only on the $T$ field 
have been identified \cite{Becker:2002nn, Balasubramanian:2004uy, vonGersdorff:2005bf, Berg:2005yu,Conlon}.

When two moduli fields are kept in the effective theory, the situation changes and new possibilities arise. 
For instance, if we consider a low-energy effective theory with the dilaton $S$ and the overall K\"ahler 
modulus $T$ we have that $n_S + n_T = 4$, and the necessary condition (\ref{necessstring}) is therefore 
comfortably satisfied. This means that, in principle, for a suitable superpotential it is possible to find 
non-supersymmetric Minkowski minima\footnote{See ref.~\cite{Brustein:2004xn} for an example.}.  
However, the $T$ field cannot be much heavier than the $S$ field, and must also contribute in a significant 
way to supersymmetry breaking, since the $S$ field does not lead to a viable situation in the limit in which 
the effect of $T$ is negligible. On the other hand, the converse situation where the $S$ field has a small 
impact compared to the $T$ field, can be compatible with stability. More precisely, one can use eqs.~(\ref{thetacrit}) 
to infer that the Goldstino angle $\theta$ is in this case constrained to be in the interval $\theta\in\left[\pi/4,\pi/2\right]$. 
This implies that:
\be
\frac{|F_T|/{\rm Re}\,T}{|F_S|/{\rm Re}\, S} >  \sqrt{3}\,.
\label{FTFS}
\ee
The result (\ref{FTFS}) implies in particular that the contribution to soft scalar masses for the matter fields coming 
from the dilaton, which has the nice feature of being approximately flavor-universal \cite{ln}, tends to be smaller
than the one coming from the K\"ahler modulus, which is instead generically non-universal.

When more than two moduli fields are involved, the situation remains similar to the two-field case. 
In order to achieve stability, one needs that the fields with sizeable auxiliary fields should have 
inverse curvatures that add up to more than $3$. This puts relevant constraints also on the type of models 
discussed in ref.~\cite{Ferrara:1994kg,Derendinger:2005ed}, where the first derivatives $W_r$ with 
respect to some of the moduli $\Phi_r$ vanish at the vacuum. Since $W \neq 0$, this implies that these 
fields break supersymmetry. Their contribution to the value of the scalar potential $V$ at the minimum
contains a term proportional to $(\sum_r n_r) |W|^2$, which overcomes the negative term $-3 |W|^2$
if $\sum_r n_r > 3$. In that case, the cosmological constant is automatically positive. However, $V$
is in general not positive definite, since $W_r \neq 0$ away from the vacuum, and 
stability is therefore still an issue, although the necessary condition on the curvatures is satisfied. 
In the special no-scale subcase of this situation in which $W$ does not depend at all on 
the superfields $\Phi_r$, $V$ becomes semi-positive definite, and stability is guaranteed 
\cite{noscale}. However, in this case $V$ does not depend at all on the pseudoscalar 
axions belonging to $\Phi_r$, and these therefore have a vanishing mass. 

An interesting deformation of the situation described by eqs.~(\ref{stringk}) can be obtained by considering
warped geometries. The simplest case where such a possibility is realized and becomes extremely relevant
is the supersymmetric generalization of the five-dimensional Randall--Sundrum scenario \cite{RS}.
In that simplest case, there is a single modulus $T$, controlling the size of the internal dimension, and the
K\"ahler potential of the effective theory has the form $K= - 3\, {\rm ln} [(1- e^{-k(T + T^\dagger)})/k]$,
where the dimensionful parameter $k$ characterizes the AdS curvature \cite{RSeffectiveK1,RSeffectiveK2}.
The K\"ahler curvature is easily computed and turns out to be constant and independent of $k$: $R_T=1/3$. 
This means that the situation is identical to the flat case\footnote{This can be understood from the fact that
the effect of the warping can be completely eliminated through a holomorphic field redefinition plus
a K\"ahler transformation, which both leave the curvature invariant.} with $n_T=3$, which is marginally
excluded by our necessary condition for local stability (\ref{necessstring}).

\section{Uplifting}
\setcounter{equation}{0}

It is interesting to see how the idea of obtaining a non-supersymmetric Minkowski or dS vacuum 
by uplifting a supersymmetric AdS vacuum fits into our study. This idea was recently proposed in ref.~\cite{KKLT} 
in the context string/M-theory compactifications, exploiting the fact that the superpotentials generated by 
background fluxes \cite{GVW} and by non-perturbative effects like gaugino condensation \cite{DW,JP}
may generate a scalar potential fixing some or even all the geometric moduli of the compactification.
In the context of compactifications to four dimensions of type IIB string theory, it has been shown 
in ref.~\cite{GKP} that background fluxes stabilize all the complex 
structure moduli as well as the dilaton. In models with just one K\"ahler modulus, it was shown in 
ref.~\cite{KKLT} that non-perturbative effects could be used to stabilize the remaining K\"ahler modulus 
at a supersymmetric AdS vacuum. This vacuum could then be uplifted to a non-supersymmetric 
Minkowski/dS vacuum by breaking explicitly supersymmetry through the introduction of anti-branes 
located in a region with strong red-shift, whose net effect is the addition of a positive term in the effective
scalar potential. This interesting uplifting mechanism might be realized also in other ways, for example 
involving vector multiplets \cite{Duplift, choifalk,vz,beatriz}. Actually, from a low-energy effective field theory point 
of view, it can in principle be implemented by using as uplifting sector any kind of theory leading to 
spontaneous supersymmetry breaking, provided the supersymmetric sector is appropriately shielded
from this uplifting sector.

In principle it should be possible to realize the idea of uplifting within the setup we have considered here, 
as the only assumptions made were that supersymmetry breaking is dominated by chiral superfields 
and that these fields have negligible mixings in the effective K\"ahler potential.
These two restrictions do not seem incompatible with the idea of uplifting. To clarify this let us consider 
a theory with $m$ chiral superfields $\Phi_\alpha$ (with $\alpha = 1,\dots\,m$) defining the 
sector that would lead to the supersymmetric AdS vacuum, and $n-m$ fields $\hat \Phi_r$ (with 
$r = 1,\dots\,n-m$) defining the ``uplifting sector". Let us also assume that the K\"ahler potential and 
the superpotential both split into two distinct pieces associated with these two sectors: 
$K_{\rm tot} = K + \hat K$ and $W_{\rm tot} = W + \hat W$. Due to gravitational effects, the two sectors 
will unavoidably interact and influence each other. In general, the structure of the potential for the whole 
theory will thus be completely different from the sum of the potentials coming from the two sectors 
if computed independently. Nevertheless there are particular circumstances under 
which the uplifting sector has a mild effect on the supersymmetric sector, thereby justifying its name and 
leading to an interesting situation.

One simple possibility to realize such a situation, that we would like to emphasize in order to illustrate
the point, is that the uplifting sector is a theory where supersymmetry is spontaneously broken 
independently of gravitational effects at a scale $M_{\rm break}$ that is much lower that the Planck 
mass $M_{\rm P}$ but still much larger than the gravitino mass $m_{3/2}$. Technically, this means 
that for that sector of the theory, all the dimensionful quantities are small compared to the Planck scale. 
Restoring the explicit dependence on $\kappa = M_{\rm P}^{-1}$ and proceeding along the same 
lines as in ref.~\cite{Weinberg:1982id}, one finds then that the scalar potential of the whole effective 
theory is just given by
\bea
\begin{array}{lll}
V_{\rm tot} \!\!\!&\simeq&\!\!\! \displaystyle{
e^{\kappa^2 K} \Bigg[\sum_{\alpha,\beta=1}^{m} K_{\alpha \bar \beta}^{-1}\,
\Big(W_\alpha + \kappa^2\, K_\alpha \, W \Big)\,
\Big(\bar W_{\bar \beta} + \kappa^2\, K_{\bar \beta} \, \bar W\Big)\,
- 3\,\kappa^2\, |W|^2\Bigg]} \smallskip\ \\
\!\!\!&\;&\!\!\! \displaystyle{+\,e^{\kappa^2 K} \Bigg[
\sum_{r,s=m+1}^n \hspace{-5pt} \hat K_{r \bar s}^{-1}\,
\hat W_r \, \bar {\hat W}_{\bar s} \Bigg]}\smallskip\ \\
\!\!\!&\equiv&\!\!\! V_{\rm susy} + V_{\rm uplift} \,.
\end{array}
\label{potabove}
\eea
Since the masses of the scalar fields belonging to the uplifting sector are much larger than those of
the supersymmetric sector, they can be integrated out. This means that the potential (\ref{potabove})
can be evaluated with the fields $\hat \phi_r$ frozen at their vacuum expectation values. 
The last bracket reduces then to a positive constant equal to $\hat K_{r s}^{-1} \,\hat F_r \,\bar {\hat F}_{\bar s} 
\sim M_{\rm break}^4$, where $M_{\rm break}$ is the supersymmetry breaking scale in the uplifting 
sector. The presence of the uplifting sector therefore results in the addition of a term $V_{\rm uplift}$ 
to the potential $V_{\rm susy}$ of the supersymmetric sector that has a mild dependence on the 
fields through the factor $e^{\kappa^2 K}$, 
which is fixed once the supersymmetric sector has been specified\footnote{It is possible to construct 
similar models that lead to an uplifting potential with a functional form that is not directly linked to $K$. 
One possibility is for example to introduce a vector multiplet gauging some isometry in the uplifting sector,
with a generic gauge kinetic function $f$ depending on the chiral superfields of the supersymmetric sector. 
The corresponding D-term potential gives then a positive contribution to the uplifting potential 
with a functional dependence on the fields that is proportional to $f$.}.
The net effect of this term is to shift the vacuum expectation values of 
the fields in the supersymmetric sector and give a positive contribution of order $M_{\rm break}^4$ to the total 
potential at the vacuum. This contribution can be tuned to cancel the negative contribution of order 
$m_{3/2}^2 M_{\rm P}^2$ coming from the supersymmetric sector, by choosing 
$M_{\rm break} \sim \sqrt{m_{3/2} M_{\rm P}}$, which is compatible with the assumption that
$m_{3/2} \ll M_{\rm break} \ll M_{\rm P}$. Simple models of this type were constructed for instance
in ref.~\cite{LSuplift,RSeffectiveK2} (see also \cite{b2b,b2bwarped})\footnote{Note that in these
constructions, it is $e^{-K/3}$ and not $K$ itself that is separable. This is however not important for
the discussion presented in this section.}.

This example clearly shows that models based on the idea of uplifting can be thought as ordinary
models involving a larger set of degrees of freedom, which includes in particular those of the uplifting
sector. The potential $V_{\rm uplift}$ can be interpreted as the remnant of the uplifting sector 
after spontaneous supersymmetry breaking has occurred and the involved degrees of freedom
have been integrated out in that sector. In other words, this reasoning means that the non-supersymmetric 
effective Lagrangian ${\cal L}_{\rm susy} + V_{\rm upllift}$ can be made supersymmetric by 
integrating in the heavy fields realizing non-linearly supersymmetry in the uplifting 
sector\footnote{The main qualitative difference between the examples presented
here and the model of ref.~\cite{KKLT} lies in the way the two sectors are shielded 
from each other. Here the crucial point is that the breaking scale is much smaller than $M_{\rm P}$,
whereas in ref.~\cite{KKLT} the breaking scale is high but its effects are red-shifted by the warping 
of the geometry.}.
When interpreted in this way, and provided that they involve only chiral multiplets, uplifted models are
subject to the necessary conditions for stability that we have derived in the previous sections. 
But as usual, even when these necessary conditions are fulfilled, the question of whether or not the 
extremum is a true minimum depends on the details of the theory. 
Since supersymmetric AdS vacua can be saddle points, when they are uplifted they can give rise to instabilities. The interesting 
question of whether the resulting non-supersymmetric Minkowski/dS vacuum is eventually a stable minimum 
or an unstable saddle point must then be studied case by case (see for example \cite{Choi:2004sx,Lust:2005dy}). 

\section{Conclusions}
\setcounter{equation}{0}

In this paper we have studied the issue of stability in the context of minimal supergravity theories with $n$
chiral superfields. We have studied under what circumstances one can have non-supersymmetric Minkowski
minima of the scalar potential. Although a general analysis of the  eigenvalues of the Hessian matrix is substantially
involved, we have found a remarkably simple necessary condition that only involves the K\"ahler potential and 
is independent of the exact form of the superpotential of the theory, as well as important restrictions on the 
Goldstino direction, again depending only on the K\"ahler curvature. These conditions must be fulfilled in order to 
have the possibility of finding non-supersymmetric Minkowski minima in the low energy theory. Nevertheless, 
they are necessary but not sufficient conditions, and only the exact form of the superpotential (as well as the K\"ahler
potential) will determine if the vacuum is really a minimum or not.

These results are relevant in several respects. From the effective theory point of view, they can give interesting information 
about soft terms, as they restrict the relative sizes of the auxiliary fields which is of importance for model building. 
In the context of string models, they should also be very useful in the task of identifying promising models
where all the moduli are stabilized at a non supersymmetric Minkowski minimum. 
Finally, it would be interesting to extend the study performed here to include vector superfields as well,
including in particular the possibility that these vector superfields gauge some isometries of the scalar manifold. 
We leave this for future work.

\section*{Acknowledgments}

We thank G. Dall'Agata, J.-P.~Derendinger, E. Dudas, S. Ferrara, N. Prezas, R. Rattazzi, P. Severa and F. Zwirner 
for useful discussions. This work has been partly supported by the Swiss National Science Foundation and by the 
European Commission under contracts MRTN-CT-2004-005104. We also thank the Theory 
Division of CERN for hospitality.

\end{document}